\documentstyle[12pt]{article}
\newcommand{\be}{\begin{equation}}
\newcommand{\ee}{\end{equation}}
\newcommand{\ba}{\begin{eqnarray}}
\newcommand{\ea}{\end{eqnarray}}
\newcommand{\dr}{\partial}
\newcommand{\Fd}{F_{\mu\nu}}
\newcommand{\Fu}{F^{\mu\nu}}
\newcommand{\Ftd}{\widetilde F_{\mu\nu}}
\newcommand{\Ftu}{\widetilde F^{\mu\nu}}

\begin{document}
\begin{flushright}
DFUB/94 - 10\\
May 1994
\end{flushright}

\vspace{2cm}
\begin{center}
{\Large\bf
Lorentz symmetry breaking
in Abelian vector-field models\\  with Wess-Zumino interaction}

\vspace{1cm}
{\bf A. A. Andrianov}\footnote{E-mail:  andrianov1@phim.niif.spb.su}\\
{\it Department of Theoretical
Physics, University of Sankt-Petersburg, 198904 Sankt-Petersburg, Russia}

\vspace{2mm}
 and\\

\vspace{2mm}
{\bf R. Soldati} \footnote{E-mail: soldati@bo.infn.it}\\
{\it Dipartimento di Fisica "A. Righi", Universit\'a
di Bologna and Istituto Nazionale\\
di Fisica Nucleare, Sezione di Bologna,
40126 Bologna, Italy}
\end{center}

\vspace{1cm}
\begin{abstract}
We consider the abelian vector-field models in the presence
of the Wess-Zumino
interaction with the pseudoscalar matter. The occurence of the dynamic
breaking of Lorentz symmetry at classical and one-loop level is
described for massless and massive vector fields. This phenomenon appears to
be the non-perturbative counterpart of the perturbative
renormalizability and/or unitarity breaking in the chiral gauge theories.
\end{abstract}

\newpage
\section*{\normalsize\bf 1.\quad Introduction}
\hspace*{3ex}
The abelian vector-field models we consider are
given by the lagrangian which contains the Wess-Zumino interaction
(in the Minkowski space-time):
\ba
{\cal L}_{WZ} & =& -\, \frac{1}{4}\, \Fd \Fu\, +\,
\frac{1}{2}\, m^2 A_{\mu} A^{\mu}
\,-\, b\, \dr_{\mu}\, A^{\mu}\nonumber\\
&&\pm\frac{\beta^2}{2}\, \dr_{\mu} \theta \dr^{\mu}\theta \,+\,
\frac{\kappa}{4M}\, \theta\,
\Fd \Ftu,  \label{basic}
\ea
where $\Ftu \equiv 1/2 \,\,\epsilon^{\mu\nu\rho\sigma} F_{\rho\sigma}$,
the universal dimensional scale $M$ is introduced,
$m \equiv \beta_V M $ is the mass of the vector field and
the coupling parameters $\beta $ and $\kappa$ take
particular values depending on their origin as we explain below.

The Wess-Zumino interaction (the last term in (\ref{basic})) can
be equivalently represented in the following form,
\be
\int d^4x\frac{\kappa}{4M}\, \theta\, \Fd \Ftu \,=\,
- \,\int d^4x\frac{\kappa}{2M}\, \dr_{\mu}\theta\, A_{\nu} \Ftu,
 \label{top}
\ee
when it is treated in the action. Therefore the pseudoscalar field
is involved into the dynamics only through its gradient
$\dr_{\mu}\theta(x)$ due to topological triviality of abelian
vector fields.

These models
have different roots.

\noindent
{\bf 1.}\quad They may represent the anomalous part of the chiral abelian
theory when the lagrangian is
prepared in the gauge invariant form by means of integration
over gauge group$^1$ and after the Landau gauge-fixing,
\ba
{\cal L}_{ch} & =& -\, \frac{1}{4}\, \Fd \Fu\,+\,\frac{1}{2}\,
m^2 (A_{\mu} + \dr_{\mu} \vartheta)
(A^{\mu} + \dr^{\mu} \vartheta)\nonumber\\
 &&-\, b \,\dr_{\mu}\, A^{\mu}
\,+\,  \bar\psi ( \not\!\dr \,+\, i e\, ( \not\!\! A \,+ \,
\not\!\dr\vartheta) P_L) \psi,
\label{mod1}
\ea
where $P_L = (1 + \gamma_5)/2$.
In this case $\kappa \equiv e^3/12\pi^2$ ,  $\theta = M \vartheta$,
 and $\beta = m/M$. The latter relation
provides the cancellation of the ghost pole in the vector-field propagator
coupled to the chiral fermion current or, in other words, it
leads to the Proca propagator for vector field.  As it is known$^2$
this is just the anomalous vertex,
\be
\theta\, \dr_{\mu} J^{\mu}_L\, =\, i \frac{e^3}{48\pi^2}\, \theta
\,\Fd \Ftu
\ee
which yields the violation of perturbative unitarity at high energies.
Therefore the nonperturbative properties of the model (1) might be crucial
in understanding of what happens with the chiral abelian model due to breakdown
of perturbative unitarity and of power-counting renormalizability.

Indeed as it follows from canonical rescaling arguments
\ba
x \rightarrow \lambda x;\quad
\dr_{\mu} \rightarrow \dr_{\mu}/\lambda;\quad A_{\mu}
\rightarrow A_{\mu}/\lambda;\quad \theta \rightarrow \theta/\lambda;\quad
b \rightarrow b/\lambda^2;\nonumber\\ \psi \rightarrow \psi/\lambda^{3/2};\quad
S_{kin} \,+\, S_{int}^{\bot }\rightarrow S_{kin} \,+\, S_{int}^{\bot};\quad
S_{int}^{\|} \rightarrow S_{int}^{\|}/\lambda,
\ea
for $\lambda << 1$ the notion of high energies is roughly
related to the
limit $M \rightarrow 0$ or
$m =\beta_V M \rightarrow 0;\, \kappa/M \rightarrow \infty$.
Thus one expects
that the effective coupling  is rapidly increasing with
energies and therefore approaching to the
strong-coupling regime.

\noindent
{\bf 2.}\quad  On the other hand the model (\ref{basic})
with $\beta^2 \rightarrow - \beta^2$
can shed light on the breaking of unitarity in the
chiral abelian vector models (CAVM)$^3$
with the anomaly compensating ghost field
\be
{\cal L}^{\bot}_{ch} \,=\,  {\cal L}_{ch} \,-\, \frac{1}{2} m^2\,\dr_{\mu}\eta
\dr^{\mu}\eta - \eta\,\dr_{\mu} J^{\mu}_{L}
\label{mod2}
\ee
This model is renormalizable by power counting, exhibits  extended
gauge invariance$^3$ and
it is equivalent to the CAVM (2) with additional
constraint $\Box \theta = 0$ that leads to the non-local lagrangian
with purely transversal gauge fields.
However the presence of triangle anomaly is still troublesome since
it eventually
gives rise  to the lack of decoupling of the massless ghost pole$^4$ in the
transversal projector, in spite of the extended gauge or BRST invariance.

Thus the model (\ref{basic}) with the negative "ghost" sign of kinetic term for
scalar fields might be helpful to unravel the infrared region of transversal
CAVM.\\
\noindent
{\bf 3.}\quad The models family (\ref{basic}) can be obtained
as a result of dimensional reduction$^5$ of the
dim-5 abelian vector-field model with the Chern-Simons interaction,
\be
{\cal L}_{CS} \, =\, -\, \frac{1}{4}\, \Fd \Fu\, +\,
\frac{\kappa_5}{10}\, \epsilon^{\mu\nu\rho\sigma\lambda}
A_{\mu} \dr_{\nu} A_{\rho} \dr_{\sigma} A_{\lambda}
\,- \,b\, \dr_{\mu}\, A^{\mu}
\ee
where now $\mu,\ldots = 0,\ldots,4;\quad dim[\kappa_5] = - 3/2$
and the metric is $diag(+, -, -, -, +)$.
This choice of metric provides the
correct sign for kinetic terms of $A_{\mu}$ and $\theta$ fields in the
four-dimensional space-time  and
corresponds to the reduction of $O(3,2)$-symmetry to $O(3,1)$-symmetry.
Other choices, $O(4,1) \rightarrow O(3,1)$ or $O(4,1) \rightarrow O(4)$,
lead to the ghost field, either the scalar one or the vector one respectively.

Then on the stationary $x_4$-independent solutions of the
field equations
one can perform the dimensional reduction of this space dimension
$|x_4| \leq \tau/2 \rightarrow 0$ so that
$$ \dr_4 A_{\mu} \simeq 0;\quad A_4 \equiv
\tau^{-1/2} \beta \theta;\quad
A_{\mu}|_{\mu\not= 4} \rightarrow \tau^{-1/2} A_{\mu}.$$
The reduced dim-4 lagrangian,
$$ {\cal L}_{WZ}\, = \,\int^{\tau/2}_{-\tau/2}\! dx_4\, {\cal L}_{CS}
\simeq \tau {\cal L}_{CS}|_{\tau \rightarrow 0}$$
takes the form (1) with $m^2 = 0$ and the coupling parameter depending on the
reduction scale $ \kappa_{WZ}/M = 3\kappa_{5} \beta/ 5\sqrt\tau$.
Thereby we have
two options either to impose $\beta \sim \sqrt{\tau M}$
 or to assume that the CS coupling
is vanishing $ \kappa_{5} \sim \sqrt\tau /M$. In the first case one obtains
the model (\ref{basic}) without kinetic term for $\theta$-field while in
the second one the kinetic term survives.

Thus the properties of dim-5 CS and dim-4 WZ models are intimately connected
as within the perturbative expansion as well as beyond it.\footnote{This
connection is manifested in Eqs.of motion, their classical
solutions and furthermore in the comparison of one-loop effective potentials.}

\vspace{5mm}

Now let us take the
model (\ref{basic})  on its own right and
proceed to the evaluation of tree-level
"classical" solutions, which may be considered as a starting
point to develop the loop "quasiclassical" approximation.
\section*{\normalsize\bf 2.\quad Equations of motion and "vacuum" solutions}
\hspace*{3ex} The Euler-Lagrange equations for the model (\ref{basic}) read:
\ba
\dr_{\mu}\,\Fu\,+\, m^2 \,A^{\nu}\,+\, \dr^{\nu}b\,
- \,\frac{\kappa}{M}\,\dr_{\mu}\theta \,
\Ftu\,=\,0;\nonumber\\
\dr_{\mu}\,A^{\mu}\,=\,0\,\Rightarrow \,\Box\,b\,=\,0;\nonumber\\
\pm\,\beta^2\,\Box\,\theta\,=\,\frac{\kappa}{4M}\,
\Fd \Ftu.
\label{class}
\ea
We search for the constant solutions
for $\dr_{\mu}\,\theta$ and related field-strength solutions.
Let us classify four distinguished sectors:
\begin{description}
\item[Case A.] $m = 0,\quad \beta = 0$, i.e. the vector field is massless and
there is no kinetic term for scalar field which becomes a pure WZ field.
\item[Case B.] $m \not= 0,\quad \beta = 0$, the vector field is massive
but
the scalar one represents a pure WZ field.
\item[Case C.] $m = 0, \quad  \beta \not= 0$, vector and scalar
fields are
massless and the scalar field is a D'Alembert field or a "ghost" massless
scalar for $\beta^2 \rightarrow - \beta^2$.
\item[Case D.] $ m \not= 0;\,  \beta \not= 0$,
the vector field is massive
but the scalar one is still massless.
\end{description}
We shall see that the classical "vacuum" solutions  and the one-loop
effective potentials are different in above cases.

It is evident from Eq.(\ref{class}) that the constant solutions
\be
 \dr_{\mu} \theta\,=\, M \eta_{\mu}\,=\, const;\quad \Fd\,=\, 0
\ee
are there for all the cases. In {\bf A}, {\bf B}
the classical action vanishes
for these Lorentz symmetry breaking solutions and
therefore the Lorentz symmetric
vacuum configuration $\eta_{\mu}\,=\, 0$ is degenerate.
The quantum effective potential (in the Coleman-Weinberg spirit$^6$)
is needed to select out the true vacuum configuration (see below).
In {\bf C}, {\bf D} for $\eta_{\mu} \eta^{\mu} \not=  0$
the above solution shifts
the potential energy  by a constant which might lead to an inequivalent
quantum theory. Again  the quantum effective potential
is helpful for the resolution of the true vacuum configuration.

\medskip

Let us also analyze constant solutions with $\Fd \not= 0$
which correspond to
$A_{\nu} \,=\, 1/2\;\Fd x^{\mu}$.
{}From Eqs.(\ref{class}) it immediately follows that possible solutions
obey the condition $ \Fd \Ftu = 0$. In {\bf A}, {\bf C}
the constant field-strength configuration and the
constant value of $\dr_{\mu} b \,\equiv\, \zeta M^2 n^{(1)}_{\mu}$
are allowed.
The first Eq.(\ref{class}) now reads,
\be
 \zeta M^2 n^{(1)}_{\nu}\,=\, \kappa \eta^{\mu}\,\Ftd \label{const}
\ee
and therefore $\eta^{\mu}  n^{(1)}_{\mu} = 0$.
The arbitrary field-strength $\Fd$ can be conveniently represented in terms of
linear-independent vectors
\ba
[ \eta_{\mu} \equiv n^{(0)}_{\mu};  n^{(1)}_{\mu}; n^{(2)}_{\mu};
n^{(3)}_{\mu}] \equiv n^i_{\mu};\quad
\epsilon^{\mu \nu \rho \sigma} \eta_{\mu}  n^{(1)}_{\nu}
n^{(2)}_{\rho} n^{(3)}_{\sigma} \not= 0; \nonumber\\
\Fd \,=\, \sum_{i<j} a_{ij} n^i_{[\mu} n^j_{\nu ]};\quad
n^i_{[\mu} n^j_{\nu ]}\;\equiv\;n^i_{\mu} n^j_{\nu} \,-\, n^i_{\nu} n^j_{\mu}.
\ea
The above conditions mean that only three independent vectors are
available in building  solutions,
namely $\eta_{\mu}; n^{(2)}_{\mu}; n^{(3)}_{\mu}$.

Let us impose the complete
decoupling of $b$-field, i.e. set $\zeta
\rightarrow 0$. Then
the solution is parametrized as follows:
\be
\Fd \,=\, a \eta_{[\mu} n^{(2)}_{\nu ]} . \label{field}
\ee
If $\eta_{\mu} \eta^{\mu} \not= 0$ the vector $n^{(2)}$ can be chosen
orthogonal to $\eta^{\mu}$ yielding
\be
\Ftd \,=\, \tilde a  n^{(1)}_{[\mu} n^{(3)}_{\nu]}.
\ee
The coefficients $a, \tilde a$ are not fixed by Eqs. of motion as well as the
vector $\eta_{\mu}$.

In {\bf B}, {\bf D} the
constant solution for $\Fd$ is incompatible
with constant $\eta_{\mu}$ but rather needs the configuration
$\theta (x) = M \left(\alpha x^2/2 +  (\eta\cdot x)\right),\;
\dr_{\mu}\theta = M \left(\alpha  x_{\mu} + \eta_{\mu}\right)$.
However from Eq.(\ref{class}) one obtains the relations
\ba
\frac{m^2}{2} \Fu &\,=\,& \kappa\, \alpha \Ftu;\quad (\frac{m^4}{2}\,+\,
\kappa^2\,\alpha^2)
\Fd\,\Fu \,=\, 0;\nonumber\\
\Fu\,\Ftd &=& 0 = \pm \frac{8M^2\beta^2\alpha}{\kappa}
\ea
which lead to $\Fu = 0,\, \alpha = 0$
in {\bf D}, whereas in {\bf B} we are
lead to $ \Fd \Fu = \Fd \Ftu = 0$ which corresponds to the radiation-like
constant field $\vec E \bot \vec H,\, |\vec E| = |\vec H|$.

We conclude that, if there is a smooth massless limit for vector fields and
the kinetic term for $\theta $ field is present, then
the Lorentz symmetry breaking (LSB) may be induced by the non-trivial average
value for $\dr_{\mu}\theta(x)$, but unlikely due to constant field-strength
configurations. Therefore let us examine
the LSB induced by a non-zero pseudoscalar background
$\dr_{\mu}\theta = M \eta_{\mu} \not= 0$
 when the quantum effects are taken into account.
\section*{\normalsize\bf 3.\quad Spectrum of vector fields in
$\eta$-background}
\hspace*{3ex} It is evident from Eqs.(\ref{class}) that the longitudinal
components decouple ($\dr_{\mu}A^{\mu} = 0$) and the auxiliary field $b(x)$ is
free. In the momentum representation
the Eqs.(\ref{class}) read (in the constant $\eta$-background),
\be
\biggl\lbrace(p^2 \,-\,m^2) g^{\mu\nu}\,+
\,i\kappa  \epsilon^{\mu\nu\rho\sigma}
\eta_{\rho} p_{\sigma}\biggr\rbrace\, A_{\nu}^{\bot}\,
\equiv  \mbox{\bf K}\cdot A^{\bot} \,=\,0.
\label{mom1}
\ee
Let us denote
\be
{\cal E}^{\mu\nu}(p) \,=\, \kappa  \epsilon^{\mu\nu\rho\sigma}\eta_{\rho}
p_{\sigma} \,=\, - {\cal E}^{\mu\nu} (- p), \label{eps}
\ee
and multiply Eq.(\ref{mom1}) by a transposed matrix $ \mbox{\bf K}^t =
\mbox{\bf K}^{\ast} = \mbox{\bf K} (- p)$
which have the same eigenvalues as {\bf K} due to its hermiticity,
\be
\mbox{\bf K}^t \mbox{\bf K}\cdot A^{\bot} \,=
\biggl( (p^2 - m^2)^2 \mbox{\bf 1}\,+\,
\hat{\cal E}^2\biggr) A^{\bot} \,=\,0.
\label{mom2}
\ee
The matrix $\hat{\cal E}^2$   represents in fact the projector
on a two-dimensional plane (for $\eta_{\mu}, p_{\mu}$ not collinear),
\ba
{\cal E}_{\mu\nu} {\cal E}^{\nu\lambda} \,&=&\,\kappa^2
\biggl\lbrace \delta_{\mu}^{\lambda}
\biggl(\eta^2 p^2 \,-\,(\eta \cdot p)^2\biggr) \,-\,
(\eta^{\lambda} \eta_{\mu} p^2  +  p^{\lambda}p_{\mu} \eta^2) \,+\,
(\eta^{\lambda} p_{\mu} + p^{\lambda} \eta_{\mu}) (\eta\cdot p)\biggr\rbrace
\nonumber\\
&\equiv& \kappa^2 \biggl(\eta^2 p^2 - (\eta\cdot p)^2\biggr)
[\mbox{\bf P}_2]^{\lambda}_{\mu},\quad
\mbox{\bf P}_2^2 \,=\,\mbox{\bf P}_2;\quad \mbox{\rm tr}\mbox{\bf P}_2 = 2 .
\ea
Evidently,
\be
\hat{\cal E}\,\mbox{\bf P}_2 = \hat{\cal E},\quad \mbox{\bf P}_{\bot} \,
\mbox{\bf P}_2 = \mbox{\bf P}_2,\quad
\mbox{\bf P}_2\cdot p = \mbox{\bf P}_2 \cdot\eta = 0.
\ee
Therefore in the $\eta_{\mu}$-direction one finds the free massive field,
\be
(p^2 - m^2) A^{\mu}_{\eta} = 0;\quad A^{\mu}_{\eta} \equiv
\frac{\eta^{\mu}}{\eta^2}
(\eta\cdot A^{\bot}),
\ee
whereas in the two-dimensional plane selected by
$\mbox{\bf P}_2$ the dispersion
law is different from a free-particle one,
\be
\biggl( (p^2 - m^2)^2 \mbox{\bf I} + \hat{\cal E}^2\biggr) \mbox{\bf P}_2
\cdot A \,=\,
\biggl( (p^2 - m^2)^2 + \kappa^2 (\eta^2 p^2 - (\eta\cdot p)^2)\biggr)
\mbox{\bf P}_2 \cdot A = 0.  \label{mom3}
\ee
Let us restrict ourselves to
space-like $\eta_{\mu}$ and choose the
coordinate frame where $\eta = ( 0, \mbox{\boldmath $\eta$})$. Then the energy
spectrum is defined by the following dispersion law,
\be
p_0^2 = \mbox{\bf p}^2 + m^2  + \frac{\kappa^2
\mbox{\boldmath $\eta$}^2}{2} \,\pm\,
\sqrt{\frac{\kappa^4 (\mbox{\boldmath $\eta$}^2)^2}{4} + m^2 \kappa^2
\mbox{\boldmath $\eta$}^2 +
\kappa^2 (\mbox{\boldmath $\eta$} \mbox{\bf p})^2} \geq 0. \label{spectr}
\ee
Thus in the plane orthogonal to both $\eta_{\mu}$ and $p_{\mu}$ there appear
two
types of waves (for two different polarizations).
In the soft-momentum region ($\mbox{\bf p}^2 << m^2 + \frac{\kappa^2
\mbox{\boldmath $\eta$}^2}{4}$)
 one reveals two massive excitations with
masses,
\be
 m^2_{\pm} \approx m^2 +  \frac{\kappa^2 \mbox{\boldmath $\eta$}^2}{2} \,\pm\,
\sqrt{\frac{\kappa^4 (\mbox{\boldmath $\eta$}^2)^2}{4} + m^2 \kappa^2
\mbox{\boldmath $\eta$}^2 }.
\ee
In particular, when the bare vector particle is massless, then after the
interaction with LSB background the mass splitting
arises and in the soft-momentum region we find only one polarization for
nearly massless excitations but a complementary polarization behaves as a
massive one.
\section*{\normalsize\bf 4.\quad Effective potential for pseudoscalar field}
\hspace*{3ex} Let us examine the role of quantum corrections
in  the formation of v.e.v. for $\dr_{\mu} \theta$ and derive the one-loop
effective potential for the gradient of WZ field induced by the virtual
creation
of vector particles\footnote{Insofar as the field $\theta$ is massless
the action (\ref{basic}), (\ref{top}) is invariant under field translations
$\theta \rightarrow \theta + const$ and the conventional
effective potential for this field is irrelevant.}.
We follow the recipe of the background-field method$^{7,8}$ to
obtain the effective potential and consider the second variation of
the action $S_{WZ}$ around constant \underline{space-like}
$\dr_{\mu} \theta = M \eta_{\mu},\quad \eta^2 < 0$ and
zero vector-field configurations.

For space-like $\eta_{\mu}$
the energy spectrum of vector
fields is real (see Eq.(\ref{spectr})). Therefore one can employ the causal
prescription for vector-field propagators and furthermore perform
the Wick rotation in computing the effective action. Then
the transversal part of the (euclidean) vector-field action reads
\be
A^{\bot}_{\mu} \left[
(- \Box + m^2) \delta_{\mu\nu}
- {\cal E}_{\mu\nu} (\hat p) \right] A^{\bot}_{\nu}
= \left( A^{\bot} \cdot \mbox{\bf K} A^{\bot} \right).
\ee
The one-loop effective potential is then obtained from the functional
determinant of the above operator $\mbox{\rm Det}\mbox{\bf K} =
(\mbox{\rm Det}\mbox{\bf K}^t \mbox{\bf K})^{1/2}$ in the conventional way$^8$,
\ba
V_{eff} &\equiv& V^{(0)} + V^{(1)};\quad V^{(0)} = \pm
\frac{\beta^2_{bare} M^2}{2} \eta^2;\nonumber\\
V^{(1)} &=& \frac{1}{Vol}\biggl\lbrace\frac14\mbox{\rm Tr}
\left[\mbox{\bf P}_2 \log \mbox{\bf K}^t(\eta) \mbox{\bf K}(\eta)
\right]\biggr\rbrace \,-\, \left\{\eta_{\mu} = 0\right\}\nonumber\\
&=& \frac12 \int_{|p| < \Lambda}\frac{d^4p}{(2\pi)^4}\,
\ln \left(1 \,+\, \frac{\kappa^2 (\eta^2 p^2 - (\eta\cdot p)^2)}
{(p^2 + m^2)^2}\right) \label{effpot1}
\ea
where the relations corresponding to Eqs.(\ref{mom1}),
- (\ref{mom3}) are used
for the euclidean space metric ($\eta^2 > 0$
from now on). One can evaluate the effective
potential, for instance, by means of perturbative expansion,
\be
V^{(1)} (\eta) \,=\, \frac12 \int_{|p| < \Lambda}\frac{d^4p}{(2\pi)^4}\,
\sum^{\infty}_{n = 1}\frac{(-1)^{n+1}}{n}
\frac{(\kappa^2 (\eta^2 p^2 - (\eta\cdot p)^2))^n}
{(p^2 + m^2)^{2n}} \equiv \sum^{\infty}_{n = 1} V^{(1)}_n
\label{effpot2}
\ee
The integration over angular variables is performed with the help of
the following identity,
\be
\int_{S^{(3)}} d\Omega \,(\eta^2 p^2 \,-\, (\eta\cdot p)^2)^n \,=\,
2\pi^2 \frac{(2n +1)!}{2^{2n} n!(n+1)!} \eta^{2n} p^{2n}.
\ee
In four dimensions the first two terms in (\ref{effpot2}) are
divergent and in the finite-cutoff regularization they have the
following cutoff dependence,
\ba
V^{(1)}_1 &\,=\,& \frac{3\kappa^2}{2^7\pi^2}\left(\Lambda^2
- 2m^2\ln\frac{\Lambda^2}{m^2} + m^2\right)\, \eta^2
\,+\, O\left(\frac{m^2}{\Lambda^2}\right); \nonumber\\
 V^{(1)}_2 &\,=\,& - \frac{5\kappa^4}{2^9\pi^2}
\ln\frac{\Lambda^2}{m^2} \, (\eta^2)^2
\,+\, O\left(\frac{m^2}{\Lambda^2}\right).
\ea
Evidently,
the renormalization is required with two counterterms,
\be
\Delta V(\theta) \,=\, \frac{\Delta\beta^2}{2} \dr_{\mu}\theta
\dr^{\mu}\theta \,+\, \frac{\Delta g}{4 M^4} ( \dr_{\mu}\theta
\dr^{\mu}\theta)^2.         \label{count}
\ee
We see that the second divergence cannot be generally
cured in the minimal model (\ref{basic}) and implies the inclusion
of the dim-8 vertex  into the bare lagrangian
(\ref{basic}). The appearance of higher-dimensional vertices in the
effective lagrangian is not surprising since the model is not perturbatively
renormalizable. Thus we should specify the boundary conditions
for the effective potential which provide the minimal form of
the lagrangian (\ref{basic}) at a
particular scale (by means of the fine-tuning of
all higher-dimensional vertices to zero value).
Then the form of the effective lagrangian for
other scales will be governed by the effective renormalization-group flow$^9$
\footnote{We omit for a moment the "dipole-ghost" term with four
derivatives $(\dr^2 \theta)^2$ which however is important
in the RG-flow for such models.}.

The remaining part of $V^{(1)}$ for $n\geq 3$ is finite and
contains the following terms,
\be
V^{(1)}_n |_{n\geq 3} \,=\, (-1)^{n+1} \left(\frac{\kappa^2\eta^2}
{4m^2}\right)^n \frac{m^4(2n + 1)}{16\pi^2n(n-1)(n-2)}.
\label{effn}
\ee
It seems that for massive vector fields there exists a weak coupling
and soft-momentum regime $\kappa^2\eta^2 < 4m^2$ which is perturbatively
safe, in the sense that
the boundary condition reproducing (\ref{basic}) can be imposed
at $\eta_{\mu} = 0$.

The related effective potential is found from (\ref{effpot2}) after
summation and reads
\be
V_{ren} \,=\,\pm \frac{\mu_1^2}{2} \eta^2 + \frac{g}
{4} (\eta^2)^2
\,+\, \frac{1}{32\pi^2}\ln\left(\frac{m^2 + z}{\mu^2_2}\right)
(5z^2 + 6m^2 z + m^4),
\ee
where we set $z \equiv \kappa^2 \eta^2/4$. The constants $\mu_1,\mu_2,
g$ are fixed by boundary conditions. For instance, the soft
momentum and weak coupling normalization (at $\eta_{\mu} = 0$) is given by
\be
 \pm \mu_1^2 = \pm \beta^2 M^2 - \frac{\kappa^2 m^2}{64\pi^2};\quad
\mu^2_2 = m^2;\quad g = -\frac{\kappa^4}{256\pi^2}.
\label{norm}
\ee
However such boundary conditions do not allow the massless limit
 or the strong-coupling regime which is expected to take place
at high energies (see Sec.1). In the latter case  we will use the normalization
at the fixed scale $M$ which has the meaning of a scale
for our measurements\footnote{Thereby we simplify the RG
analysis which should support our consideration.}.

\section*{\normalsize\bf 5.\quad Dynamic breaking of the Lorentz
symmetry}
\hspace*{3ex} Let us search for
the Coleman-Weinberg$^6$ instabilities in the effective potential
of pseudoscalar field that
cause the  dynamic symmetry breaking of the Lorentz symmetry (for other
scenarios of LSB, see $^{10}$). We examine
separately the models with massless and massive vector fields.

In the first case the one-loop effective potential cannot be normalized
at zero momenta, i.e. at $\eta^2 = 0$. Instead one has to provide the
basic lagrangian at the main scale of the model $\eta^2 = M^2$
\be
V_{ren} (\eta,\, \mu = \frac{\kappa M}{2}) =
\pm \frac{\beta^2 M^2}{2} \eta^2 \,+\,
\frac{5\kappa^4}{2^9 \pi^2} (\eta^2)^2 \ln\frac{\eta^2}{M^2}.
\label{V0}
\ee
The minimum is obtained from the conditions,
\be
\frac{\dr V}{\dr \eta_{\mu}} = 2\eta_{\mu} V'(\eta^2) = 0;\quad
V'(\eta^2) \equiv \frac{d V(\eta^2)}{d(\eta^2)} \,=\,
\pm \frac{\beta^2 M^2}{2} \,+\,
\frac{5\kappa^4}{2^9 \pi^2} \eta^2 \left( 2 \ln\frac{\eta^2}{M^2} + 1
\right), \label{stat1}
\ee
and
\ba
\frac{\dr^2 V}{\dr \eta_{\mu}\dr \eta_{\nu}} \,=\,
2\delta_{\mu\nu} V'(\eta^2) \,+\, 4 \eta_{\mu}\eta_{\nu}
V''(\eta^2) \geq 0,\nonumber\\
V''(\eta^2) \equiv \frac{d^2 V(\eta^2)}{(d(\eta^2))^2} \,=\,
\frac{5\kappa^4}{2^9 \pi^2} \left( 2 \ln\frac{\eta^2}{M^2} + 3
\right), \label{stat2}
\ea
The symmetric solution $\eta_{\mu} = 0$ leads to the minimum if
$V'(0) > 0$ which corresponds to the positive sign in the first term
(\ref{V0}). In the latter case other solutions may arise
for a strong coupling $\kappa$ when
$V'(\eta^2) = 0$.

In order to find the critical value of  $\kappa$ let us
substitute this equation (\ref{stat1}) into (\ref{stat2})
to provide  $V''(\eta^2) \geq 0$. Then one obtains
that for $\kappa^4 \geq 128\pi^2 e^{3/2} \beta^2 /5$ the second minimum
appears. However it lies higher than the symmetric one as compared
with the value of the effective potential at $\eta_{\mu} = 0$.
They are degenerate when
$\kappa^4_{cr} \,=\, 256\pi^2 e \beta^2 /5$ and for higher values
of $\kappa$ the LSB vacuum is favourable. By its character the corresponding
phase transition is of the first order since at $\kappa_{cr}$ the
v.e.v. of $\eta^2$ jumps to $\eta^2_{cr} = M^2 /e$. This v.e.v. entails
the LSB due to creation of space-like constant gradient of pseudoscalar field
$\dr_{\mu}\theta \sim M^2 e^{-1/2}\,\, (0, \mbox{\bf n});\,
\mbox{\bf n}^2 = 1 $.

When going back to Sec.1 one can conclude
that the plausible scenario
of what happens in the chiral gauge model (\ref{mod1})
with Proca vector fields at high
energies is the LSB at strong coupling. This is what might be behind
the breaking of perturbative unitarity in such models.

If $\beta = 0$ (Sec.2, Case {\bf C}, the pure WZ interaction
without a kinetic term for $\theta$-field)
the LSB
still occurs and the Lorentz symmetric extremum becomes a maximum.
For the negative sign of the first term in Eq.(\ref{stat1}) the LSB minimum
always exists and  a normalization scale is not there to prevent the
Lorentz symmetric vacuum from decay.

Let us extend our analysis to the massive vector-field models.
We pay the special attention to the power-counting
renormalizable chiral model (\ref{mod2}) with
transversal vector fields. This model
is suitably described
in the "ghost" sector by the effective lagrangian(\ref{basic}),
with a "ghost" sign of the  kinetic term for $\theta$-field
($\beta^2 \rightarrow - \beta^2;\,m = \beta M$) and
can be consistently normalized
by the choice (\ref{norm}) at the infrared point.
The LSB conditions (\ref{stat1}), (\ref{stat2}) are fulfilled both in
the strong
and the weak coupling regimes and the LSB minimum is unique since
$V''(\eta^2) > 0$ for positive $\eta^2$.
 In particular,
\ba
\eta^2_{min} \,&\simeq &\, m^2 \frac{128 \pi^2}{5 \kappa^4
\ln (32 \pi^2/\kappa^2)}
\quad\mbox{for}\quad \kappa << 1;\nonumber\\
\eta^2_{min} \,&\simeq &\, m^2 \frac{32 \pi}{\kappa^3 \sqrt{7}}
\quad\mbox{for}\quad \kappa >> 1.
\ea
Thus the symmetric vacuum in such a model is always unstable.

Once the Lorentz symmetry is spontaneously broken  one should
expect the occurence of Goldstone modes in the spectrum of
fluctuations, $\dr_{\mu} \widetilde\theta = \dr_{\mu} \theta - M \eta_{\mu} $
 around the minimum. In our case it gives rise to the
degeneracy in the kinetic term of pseudoscalar fluctuations.
As it follows from (\ref{stat2}),  the
second variation contains the projector on the $\eta_{\mu} $
direction. Consequently, after Wick rotating to the Minkowski
space-time, the kinetic term $-1/2\, \left((\eta\cdot \dr)
\widetilde\theta (x)\right)^2$
describes the dynamics of a massless free mode whose support,
in the momentum space, lies on the space-like hyperplane
$\eta_{\mu} p^{\mu}$ ( this feature typically arises in the
quantization of Yang-Mills fields in algebraic non-covariant gauges$^{11}$).
In other directions the dynamics is generated by
higher-derivative "ghost-dipole" terms in the effective action,
$\sim (\dr^2 \theta)^2$.

For a different sign of kinetic term and for $\beta = 0$ there is no
LSB solutions with space-like $\eta_{\mu}$ but there are time-like
LSB configurations which yield the extremum for the one-loop action.
However the very notion of vacuum energy should be carefully
revised
since those configurations are involved
in the construction of the Hamiltonian by means
of Legendre transformation. We postpone the analysis of this problem
as well as the description of energy spectrum of $\theta$-field
fluctuations to a more detailed paper.
\section*{\normalsize\bf 6.\quad Conclusions}
As a result of the present investigation one can argue
that the cancellation of anomalies strongly prevents the chiral
theories from the Lorentz symmetry breaking.

On the other hand in the presence of Wess-Zumino interaction
the occurence of LSB, at least in small domains, seems to be natural.
Indeed, the WZ action is invariant under general coordinate
transformations, $x_{\mu} \rightarrow n_a (x_{\mu})$. This mapping
is a local diffeomorphism when
$$ \mbox{\rm det}\left[ \dr^{\mu} n_a\right] \equiv
\epsilon^{\mu\nu\rho\sigma} \dr^{\mu} n_0 \dr^{\nu} n_1
\dr^{\rho} n_2 \dr^{\sigma} n_3 \,\not=\, 0.$$
Under these transformations the fields and derivatives behave as vectors,
$$ A^{\mu} \,=\, \bar A^a (n) \frac{\dr n_a}{\dr x_{\mu}}.$$
Therefore the pseudoscalar Chern-Pontryagin density in the WZ action
turns out to be multiplied by
$det \left[ \dr^{\mu} n_a\right]$, just compensating
the change in the integration measure.
Let us select out the curvilinear coordinate system with
$\theta(x)$ as one of the coordinate vector,
i. e. $\theta \equiv \bar\theta (n) = M \eta_a n^a$, as it is always possible
inside the domains where $\dr_{\mu} \theta \not= 0 $. Then the WZ action
(\ref{top})
in new coordinates takes the form of the Chern-Simons action
in the hyperplane orthogonal to the \underline{constant} vector $\eta_a$,
\be
S_{WZ} \,=\, - \frac{\kappa}{2} \int dn_{\parallel} \eta_a
\int d^3 n^{\bot}\epsilon^{abcd} A_b \dr_c A_d,
\ee
which provides the reduction of dynamics from four to three dimensions.
Thus locally, in domains of smooth $\dr_{\mu} \theta$, one  actually
deals with the dynamics described in our paper.

We see some similarities of LSB between the
(2 +1)-dimensional
case$^{12}$, as it is connected with the dynamics in two dimensions, and
the (4 + 1)-dimensional case in its reduction to our model. These
similarities will be discussed elsewhere.

Apart from the benefit for
understanding the chiral model,
the spectral properties of the light in a pseudoscalar
medium ("pseudoscalar optics") could happen to be applicable in two situations:
first, to explore the axion matter if it exists in the Universe and,
second, to check the possibility of strong-coupling  LSB in the
neutral-pion matter under extreme conditions. If the latter one is
conceivably
characterized by a vanishing pion decay constant $F_{\pi} \rightarrow
0$ at the phase transition of chiral symmetry restoration
then  the WZ interaction of pions and
photons may be enhanced considerably to invoke LSB.
\section*{\normalsize\bf Acknowledgements}
This paper has been done starting from many fruitful discussions with Prof.
A. Bassetto to whom we express our gratitude. A.A. is indebted to
Prof. G. Nardelli for both the useful discussions and a support.
The work is partially supported by GRACENAS (Russia, Grant.No.
2040) and by Italian MURST 40\%.

\vspace{0.5cm}

\section*{\normalsize\bf References}
\begin{enumerate}
\item L. D. Faddeev, {\it Phys. Lett.} {\bf B145}, 81 (1984);
\, L. D. Faddeev and S. L. Shatashvili, {\it Theor. Math. Phys}
{\bf 60}, 206 (1986);\, {\it Phys. Lett.} {\bf B167}, 225 (1986);\\
R. Jackiw and R. Rajaraman, {\it Phys. Rev. Lett.} {\bf 54}, 1219 (1985);\\
 A. A. Andrianov and Yu. V. Novozhilov,
{\it Phys. Lett.} {\bf B163}, 189 (1985);\\
O. Babelon, F. A. Shaposhnik and C. M. Viallet,
{\it Phys. Lett.} {\bf B177}, 385 (1986);\\
K. Harada and I. Tsutsui, {\it Phys. Lett.} {\bf B183}, 311 (1987).
\item D.  Gross and R. Jackiw, {\it Phys. Rev.} {\bf D6}, 477 (1972).
\item A. A. Andrianov, A. Bassetto and R. Soldati, {\it Phys. Rev. Lett.}
{\bf 63}, 1554 (1989).
\item A. A. Andrianov, A. Bassetto and R. Soldati, {\it Phys. Rev.}
{\bf D44}, 2602 (1991); {\bf D47}, 4801 (1993).
\item G. V. Dunne and C. A. Trugenberger, {\it Ann. Phys. (N.Y.)}
{\bf 204}, 281 (1990).
\item S. Coleman and E. Weinberg, {\it Phys. Rev.} {\bf 7}, 1888 (1973).
\item B. S. DeWitt, {\it Phys. Rev.} {\bf 160}, 1113 (1967);\,{\bf 162},
1195, 1239 (1967);\\
L. F. Abbot, {\it Act. Phys. Pol.} {\bf B13}, 33  (1982).
\item I. L. Buchbinder, S. D. Odintsov and I. L. Shapiro,
{\it Effective Action in Quantum Gravity} (IOP, Bristol and Philadelphia,
1992).
\item H. Georgi, {\it Nucl. Phys.} {\bf B361}, 339 (1991).
\item H. B. Nielsen and I. Picek, {\it Nucl. Phys.} {\bf B211}, 269 (1983).
\item A. Bassetto, G. Nardelli and R. Soldati,
{\it Yang-Mills Theories in Algebraic Non-covariant Gauges}
(World Scientific, Singapore, 1991).
\item Y. Hosotani, University of Minnesota, Reports No. UMN-TH-1211/93
(1993); No. UMN-TH-1238/94 (1994).
\end{enumerate}
\end{document}